\documentclass[12pt]{article}
\usepackage{graphicx,graphics}
\usepackage{amssymb}
\usepackage{amsmath}
\usepackage{dcolumn}
\usepackage{bm}
\usepackage{color}
\def\be{\begin{equation}}
\def\ee{\end{equation}}
\newcommand{\bea}{\begin{eqnarray}}
\newcommand{\eea}{\end{eqnarray}}
\newcommand{\nn}{\nonumber}

\def\hbar#1{\backslash\hspace{-2mm}#1}

\def\nn{\nonumber}

\def\2tvec#1#2{
\left(
\begin{array}{c}
#1  \\
#2  \\
\end{array}
\right)}

\def\mat2#1#2#3#4{
\left(
\begin{array}{cc}
#1 & #2 \\
#3 & #4 \\
\end{array}
\right) }

\def\Mat3#1#2#3#4#5#6#7#8#9{
\left(
\begin{array}{ccc}
#1 & #2 & #3 \\
#4 & #5 & #6 \\
#7 & #8 & #9 \\
\end{array}
\right) }

\def\3tvec#1#2#3{
\left(
\begin{array}{c}
#1  \\
#2  \\
#3  \\
\end{array}
\right)}

\def\hbar#1{\backslash\hspace{-2mm}#1}

\def\nn{\nonumber}

\newcommand{\bt}{\begin{itemize}}
\newcommand{\et}{\end{itemize}}
\numberwithin{equation}{section}

\begin{document}

\begin{titlepage}
\begin{flushright}
P12079
\end{flushright}

\begin{center}

\vspace{1cm}
{\large\bf Dark Matters in Gauged $B-3L_i$  Model  }
\vspace{1cm}

Hiroshi Okada$^{a}$\footnote{HOkada@kias.re.kr}
\vspace{5mm}

{\it%
$^{a}${School of Physics, KIAS, Seoul 130-722, Korea}\\
}
  
  \vspace{8mm}

\abstract{We study a dark matter  model with local $B-3L_i$ symmetry that is known as anomaly free and requires a single right-handed neutrino. Here we have two dark matter candidates; that is, fermionic or bosonic one. We focus on analyzing each of the case within the light mass region, which is required by the perturbative theory of the Higgs quartic coupling. 
 }

\end{center}
\end{titlepage}

\setcounter{footnote}{0}

\section{Introduction}
Recently a single right-handed neutrino scenario  (but not  with a single $\eta$) was proposed by D. Hehn and A. Ibarra group \cite{Hehn:2012kz} and discussed promising dark matter (DM) candidates in the framework of radiative seesaw model \cite{Ma:2006km}. One of the promising achievement of a single right-handed neutrino is to introduce gauged $B-3L_i$ proposed by E. Ma \cite{Ma:1997nq, Ma:1998dp, Ma:1998dr}.

Combining appropriately the radiative seesaw model and gauged $B-3L_i$ model, in our paper,  we try to analyze two DM candidates ; that is, fermionic one (right-handed neutrino) or bosonic one (neutral inert Higgs).
The neutrino sector can be mainly led in the five-dimensional operator without any right-handed neutrinos. One also finds that these DMs can be independently analyzed from the form of the lepton mass matrix because DMs do not affect to the matrix, which is unlike the typical radiative seesaw model.

This paper is organized as follows. In Section 2, we set up our model and explain the lepton sector and Higgs sector.
In Section 3, we discuss the DM properties in the fermionic case and the bosonic case.
In Section 4, we  conclude the paper.

\section{The Model}

\begin{table}[thbp]
\centering {\fontsize{10}{12}
\begin{tabular}{||c|c|c|c|c|c|c|c||c|c|c||}
\hline\hline ~~Particle~~ & ~~$L_e$~~ & ~~$L_\mu$ ~~ & ~~ $L_\tau $~~ & ~~$e^c$~~ & ~~$\mu^c$ ~~ & ~~ $\tau^c$~~ & ~~ $N^c$~~
 & ~~$\Phi~~ $& ~~$\eta~~ $ & $\chi $\\\hline
$SU(2)_L$&$\bm{2}$ & $\bm{2}$ & $\bm{2}$ & $\bm{1}$ & $\bm{1}$ &
$\bm{1}$ & $\bm{1}$ & $\bm{2}$ & $\bm{2}$ & $\bm{1}$\\\hline
$Y_{B-3L_e}$ & $3$ & $0$ & $0$ & $-3$ & $0$ & $0$ & $3$ & $0$ & $0$ & $-3$\\\hline
$\mathbb{Z}_2$ & $+$  & $+$& $+$ & $+$ & $+$ & $+$ & $-$ & $+$ & $-$ & $+$\\
\hline
\end{tabular}%
} \caption{The field content and the charges. Here we take $i=e$ of  $B-3L_i$ as an example. } \label{tab:b-3l}
\end{table}
The seesaw mechanism can be realized by introducing the $B-3L_i$ symmetry which is
spontaneously broken at TeV scale. 
The field content of our model is shown in Table \ref{tab:b-3l} and the Lagrangian in the lepton sector is
\begin{eqnarray}
{\cal L}&=&
y_e L_e\Phi e^c + y_\mu L_\mu\Phi \mu^c  + y_\tau L_\tau\Phi \tau^c  + y'_{\mu\tau} L_\mu\Phi\tau^c  + y_{\mu\tau} L_\tau\Phi \mu^c \\
&+& y'_{e\mu} L_e\Phi \mu^c\chi + y'_{e\tau} L_e\Phi \tau^c\chi+ y_{e\mu} L_\mu\Phi e^c\chi^\dag + y_{e\tau} L_\tau\Phi e^c\chi^\dag\\
&+&
y_\nu  L_e\eta N^c + \frac{y_N}{\Lambda}\chi^2 N^cN^c  \\
&+& 
{\lambda^\nu_1 \over \Lambda} (L_\mu\phi)(L_\tau\phi) + {\lambda^\nu_2 \over \Lambda} (L_\mu\phi)^2 + {\lambda^\nu_3 \over \Lambda} (L_\tau\phi)^2
\\
&+&
 {\lambda^\nu_4 \over \Lambda^2} (L_e\phi)(L_\mu\phi)\chi + 
{\lambda^\nu_5 \over \Lambda^2} (L_\tau\phi)(L_e\phi)\chi
\\
&+&{\lambda^\nu_6 \over \Lambda^3} (L_e\phi)^2\chi^2 +
 \mathrm{h.c.}, 
\label{eq:lagrangian}
\end{eqnarray}
where $\Lambda$ is a cut-off scale to be ${\cal O}(10^{7\sim8})$ GeV to maintain $\lambda^\nu_{i}\le{\cal O}(1)$. As a result
we obtain the right-handed mass range as follows; ${\cal O}$(1) GeV $\le m_{N^C}\le$ ${\cal O}$(10) GeV, where $m_{N^C}=y_Nv'^2/\Lambda$ and $v'$ is set to be $12$ TeV to satisfy the LEP bound \cite{carena}.

After the symmetry breaking, that is $\chi=(\chi^0+v')/\sqrt2$,
$\Phi=(\phi^+,\phi)^T$  with $\phi=(\phi^0+v)/\sqrt2$ and $v=246\
{\rm GeV}$, the neutrino sector in the flavor basis can be written as
\begin{equation}
{\cal M}_\ell\simeq\left(
\begin{array}{ccc}
y_ev & \frac{y'_{e\mu}}{\Lambda}vv' & \frac{y'_{e\tau}}{\Lambda}vv' \\
\frac{y_{e\mu}}{\Lambda}vv' & y_{\mu}v & y'_{\mu\tau}v\\
\frac{y_{e\tau}}{\Lambda}vv' & y_{\mu\tau}v & y_{\tau}v
\end{array}
\right),\quad
{\cal M}_\nu \simeq
\left(
\begin{array}{ccc}
\delta m_{\nu}+\frac{\lambda^\nu_6}{\Lambda^3}v^2v'^2 &\frac{\lambda^\nu_4}{\Lambda^2}v^2v' & \frac{\lambda^\nu_5}{\Lambda^2}v^2v'\\
\frac{\lambda^\nu_4}{\Lambda^2}v^2v' & \frac{\lambda^\nu_2}{\Lambda}v^2 &  \frac{\lambda^\nu_1}{\Lambda}v^2\\
\frac{\lambda^\nu_5}{\Lambda^2}v^2v' &   \frac{\lambda^\nu_1}{\Lambda}v^2 &  \frac{\lambda^\nu_3}{\Lambda}v^2
\end{array}
\right),\label{chgd-mass}
\end{equation}
where  $\delta m_{\nu}$ is obtained through the radiative seesaw mechanism in Ref. \cite{Ma:2006km} \footnote{Here we take $i=e$ of  $B-3L_i$ as an example, however the other choices are possible because they do not change the result of the lepton mass and mixing by introducing an appropriate rotating matrix. Only the difference is the final state of the DM annihilation.}.
Here notice that $\eta\equiv(\eta^+,(\eta_R+i\eta_I)/\sqrt2)^T$ with ($\eta^0=(\eta_R+i\eta_I)/\sqrt2)$) does not
have any vacuum expectation values (VEV). Apparently these mass matrices can successfully lead the observed neutrino mass and mixing \cite{t2k, daya, reno}, and our DM candidate $N^c$ or $\eta^0$ is independent of the form of the these matrices. Hence we are apart from the discussion of lepton sector. Notice furthermore that there is no constraint of $N^c$, that is DM, from LFV violation, since the first charged lepton flavor eigenstate $L_e$ is almost  the same as the  mass eigenstate at the leading order as can be seen in Eq. (\ref{chgd-mass}).

\subsection{Higgs Sector}
The Higgs potential of this model is given by \footnote{In details, see Ref. \cite{Okada:2012np}. }
\begin{eqnarray}
V \!\!\!&=&\!\!\!
 m_1^{2} \Phi^\dagger \Phi + m_2^{2} \eta^\dagger \eta  + m_3^{2} \chi^\dagger \chi +
\lambda_1 (\Phi^\dagger \Phi)^{2} + \lambda_2 
(\eta^\dagger \eta)^{2} + \lambda_3 (\Phi^\dagger \Phi)(\eta^\dagger \eta) 
+ \lambda_4 (\Phi^\dagger \eta)(\eta^\dagger \Phi)
\nonumber \\ &&\!\!\!
+
\lambda_5 [(\Phi^\dagger \eta)^{2} + \mathrm{h.c.}]+
\lambda_6 (\chi^\dagger \chi)^{2} + \lambda_7  (\chi^\dagger \chi)
(\Phi^\dagger \Phi)  + \lambda_8  (\chi^\dagger \chi) (\eta^\dagger \eta),
\end{eqnarray}
where $\lambda_5$ has been chosen to be real without any loss of
generality.
Inserting the tadpole conditions; $m^2_1=-\lambda_1v^2-\lambda_7v'^2/2$
and $m^2_3=-\lambda_6v'^2 - \lambda_7v^2/2$, 
the resulting mass matrice are given by
\begin{eqnarray}
m^{2} (\phi^{0},\chi^0) = \left(%
\begin{array}{cc}
  2\lambda_1v^2 & \lambda_7vv' \\
  \lambda_7vv' & 2\lambda_6v'^2 \\
\end{array}%
\right) &=& \left(\begin{array}{cc} \cos\alpha & \sin\alpha \\ -\sin\alpha & \cos\alpha \end{array}\right)
\left(\begin{array}{cc} m^2_{h} & 0 \\ 0 & m^2_{H}  \end{array}\right)
\left(\begin{array}{cc} \cos\alpha & -\sin\alpha \\ \sin\alpha & \cos\alpha \end{array}\right), \nn\\\\ 
m^{2} (\eta^{\pm}) &=& m_2^{2} + \frac12 \lambda_3 v^{2} + \frac12 \lambda_8 v'^{2}, \\ 
m^2_R\equiv m^{2} ( Re \eta^{0}) &=& m_2^{2} + \frac12 \lambda_8 v'^{2}
 + \frac12 (\lambda_3 + \lambda_4 + 2\lambda_5) v^{2}, \\ 
m^2_I\equiv m^{2} ( Im \eta^{0}) &=& m_2^{2} + \frac12 \lambda_8 v'^{2}
 + \frac12 (\lambda_3 + \lambda_4 - 2\lambda_5) v^{2},
\end{eqnarray}
where $h$ implies standard model (SM)-like Higgs and $H$ is an additional Higgs mass
eigenstate. Here we identify $m_R$ as DM mass ($m_{\rm DM}$). 
The masses of $\phi^0$ and $\chi^0$ are rewritten in terms of the mass eigenstates of $h$ and $H$
as
\begin{eqnarray}
\phi^0 &=& h\cos\alpha + H\sin\alpha, \nn\\
\chi^0 &=&- h\sin\alpha + H\cos\alpha.
\label{eq:mass_weak}
\end{eqnarray}

\section{Dark Matter}
In our model, there are two DM candidates; that is, the right-handed neutrino $N^C$ or the neutral inert Higgs boson $\eta_R/\eta_I$. In bosonic case, however, we assume $\eta_R$ is DM. The mass $m_N^c$ is given by $y_Nv'^2/\Lambda$, as can be seen in the previous section, and the mass range should be ${\cal O}$(1) GeV $\le m_{\chi}\le$ ${\cal O}$(10) GeV in order to maintain $\lambda^\nu_{i}\le{\cal O}(1)$.
Hereafter, we symbolize DM as $\chi$  and strict ourselves the DM mass region of $1-10$ GeV for both cases.
Below we analyze each of DM candidate.

\subsection{Fermionic Dark Matter}
In case of fermionic DM($N^C\equiv \chi$), the effective Lagrangian is given by
 \begin{eqnarray}
{\cal L}\simeq
y_\nu(\ell_e\eta^+-\nu_e \eta^o) \chi + \frac{m_\chi}{\sqrt2v'} \chi\chi (-h\sin\alpha+H\cos\alpha) 
+ \frac{y_f}{\sqrt2}(h\cos\alpha+H\sin\alpha)f\bar f,
 \end{eqnarray}
where the first term plays an  role in explaining the observed relic density reported by WMAP \cite{wmap}.
The other terms are relevant to direct detection experiments \footnote{Notice that  our $\chi$ has a coupling of $\chi-\chi-Z'$, where $Z'$ is an extra neutral gauge boson. However the process vanishes in the spin independent scattering since $\chi$ is the Majorana type particle. }.

The annihilation cross section to explain WMAP is obtained through the $t$ and $u$ channels with the final states of an $e^--e^+$ pair and a $\nu_e-\bar\nu_e$ pair as follows \cite{d6-rad}:
\be
\sigma v_{ eff}\simeq \frac{|y_\nu|^2}{24\pi}\frac{m^2_\chi(m^4_\chi+m^4_\eta)}{(m^2_\chi+m^2_\eta)^4}v^2_{ eff},
\label{f-sigma}
\ee
where $v_{eff}$ is the relative velocity and we assume that these masses of the final states are taken to be zero.

In the direct detection, the spin independent elastic cross section $\sigma_{SI}$ with nucleon $N$ is given by 
\be
\sigma_{SI}\simeq \frac{\mu^2_{\rm DM}}{\pi}\left(\frac{1}{m^2_h}-\frac{1}{m^2_H}\right)^2
\left(\frac{m_\chi m_N\sin\alpha\cos\alpha}{vv'}\sum_q f^p_q \right)^2,
\ee 
where $\mu_{\rm DM}=\left(m_\chi^{-1}+m_N^{-1}\right)^{-1}$ is the
DM-nucleon reduced mass.
The parameters $f_q^p$ are determined by the pion-nucleon sigma term as follows:
\begin{eqnarray}
&&f_u^p=0.023,\quad
f_d^p=0.032,\quad
f_s^p=0.020,
\end{eqnarray}
for the light quarks and $f_Q^p=2/27\left(1-\sum_{q\leq 3}f_q^p\right)$ for
the heavy quarks $Q(=c,b,t)$ \cite{pdg} where $q\leq3$ implies the summation of the light
quarks. 
As a result, we obtain the elastic scattering cross section as
 \be
1.0\times10^{-52}({\rm cm^2})\lesssim\sigma_{SI}\lesssim 1.0\times10^{-50}({\rm cm^2}).
\ee

In the view of indirect detections, we should consider a constraint of no excess of the antiproton flux, which is proportional to $\sigma v_{eff.}/m^2_\chi$, reported by PAMELA \cite{Adriani:2010rc}, since our $m_\chi$ is tiny and couples to quarks. 
Moreover if the $\sigma v_{eff.}$ has $s$-wave, the amount of the cross section is retained to the current universe with no suppression of the relative velocity \cite{Okada:2012cc}. As a result antiproton might be overproduced, depending on the profile \cite{Evoli:2011id}.
In other words, our DM is completely safe, because our $\sigma v_{eff.}$ does not have $s$-wave in Eq.(\ref{f-sigma}).

The LHC experiment recently reported an invisible decay of the SM-like Higgs that the branching ratio ${\rm Br}_{inv}$ is excluded to the region $0.4\le{\rm Br}_{inv}$ \cite{Giardino:2012dp}, when $m_\chi<m_h/2$.
 Our invisible decay width is given by
\bea
\Gamma(h\rightarrow2\mathrm{\chi})
\simeq
\frac{m_h}{2\pi}\left(\frac{ m_\chi\sin\alpha}{v'}\right)^2 \sqrt{\frac14-\left(\frac{m_\chi}{m_h}\right)^2}
\left[\frac12-\left(\frac{m_\chi}{m_h}\right)^2 \right]
,
\eea
where we fix  the mass of SM-like Higgs to be $125~\mathrm{GeV}$ \cite{Higgsd1,Higgsd2}\footnote{See the Ref. \cite{Dittmaier:2012vm} for the total branching ratio of SM Higgs.}. 
As a result, we find our ${\rm Br}_{inv}$ is at most less than $10^{-3}$, which is safe for the LHC experiment.

As a conclusion, we obtain the elastic scattering cross section as
\be
1.0\times10^{-52}({\rm cm^2})\lesssim\sigma_{SI}\lesssim 1.0\times10^{-50}({\rm cm^2})\ 
{\rm with}\ 1\ {\rm GeV}\lesssim {m_\chi}\lesssim 10\ {\rm GeV},
\ee
where we take the limit of the maximal mixing of $\sin\alpha$ for simplicity. Notice here that the above result is obtained by the direct detection , since the annihilation cross section to explain WMAP has two independent parameters $y_\nu$ and $m_\eta$  and the invisible decay constraint of LHC satisfies our model.
It implies that future experiments could be tested.

\subsection{Bosonic Dark Matter}
In case of bosonic DM($\eta_R\equiv \chi$), 
the mass $m_\chi$ should be always less than the right-handed mass $m_{N^c}(\lesssim{\cal O}(10)\ {\rm GeV})$ in order to forbid the too fast decay through term $y_\nu L_e\eta N^c$, and
there are two annihilation modes; 
$t$ and $u$ channel of $2N^c\rightarrow \eta^{0(\pm)}\rightarrow 2\nu(\ell\bar\ell)$
and $s$-channel of  $2\eta\rightarrow h/H\rightarrow f\bar f$. However since the $t$ and $u$ channel does not have $s$ wave contribution, the dominant cross section is given only through the $s$ channel as follows \cite{Kajiyama:2012xg}:
\begin{eqnarray}
\sigma v_{eff.}
&\simeq&
\sum_{f}\frac{c_fy^2_f}{2\pi}
\left|\frac{\lambda_h\cos\alpha}{4m_{\chi}^2-m_{h}^2+im_{h}\Gamma_h}
+\frac{\lambda_H\sin\alpha}{4m_{\chi}^2-m_H^2+im_{H}\Gamma_H}\right|^2,
\label{b-sigma}\\
2\lambda_h&\equiv& (\lambda_3+2\lambda_4+2\lambda_5)v \cos\alpha-\lambda_8v'\sin\alpha,\nn\\
2\lambda_H&\equiv& (\lambda_3+2\lambda_4+2\lambda_5)v \sin\alpha+\lambda_8v'\cos\alpha,\nn
\end{eqnarray}
where $m_{DM}=m_R$ is DM mass,
$y_f$ is Yukawa coupling of SM matter particle,  $\Gamma_h=4.1\times 10^{-3}$ GeV \cite{higgsdecay}, $\Gamma_H=\sum_{f}\frac{c_fy_{f}^2\cos^2\alpha}{16\pi}m_H\left(1-\frac{4m_{f}^2}{m_H^2}\right)^{3/2}$, and
the color factor $c_f$ is $3$ for quarks and $1$ for leptons.

Since our DM is less than $m_h/2$, the invisible decay constraint should be considered.
Our invisible decay width is give by
$\Gamma(h\rightarrow 2\chi)\simeq\frac{\lambda_h^2}{16\pi m_h}\sqrt{1-4m_{\chi}^2/m_h^2}$, which is the same result as in Ref. \cite{Kajiyama:2012xg}.

In the direct detection, the spin independent elastic cross section $\sigma_{SI}$ with nucleon $N$ is given by 
\be
\sigma_{SI}=\frac{\mu_{\mathrm{DM}}^2}{\pi}\frac{m_N^2}{m_{\chi}^2v^2}
\left(\frac{\lambda_h\cos\alpha}{m_h^2}+\frac{\lambda_H\sin\alpha}{m_H^2}\right)^2
\left(\sum_qf_q^p\right)^2,
\ee 
where $\mu_{\rm DM}$ and $f_q^p$ has been defined in the fermionic part.

Considering all these constraints, we find the elastic scattering cross section\footnote{See the Fig.3 in ref. \cite{Kajiyama:2012xg}} as
\be
5.0\times10^{-41}({\rm cm^2})\lesssim\sigma_{SI}\lesssim 5.0\times10^{-40}({\rm cm^2})\ 
{\rm with}\ 1\ {\rm GeV}\lesssim {m_\chi}\lesssim 10\ {\rm GeV}.
\ee
The range of the
cross section is consistent with  CRESSTII~\cite{cresst}, CoGeNT~\cite{cogent}, and DAMA~\cite{dama}.

\section{Conclusions}
We studied the Dark matter features of right-handed fermion $N^c$ and neutral inert boson $\eta_R$ in
a local $B-3L_i$ partially radiative model. Due to the perturbative theory of the Higgs quartic coupling, both of DMs masses should be within ${\cal O}$(10) GeV. In case of the fermionic DM, we find that the maximum scattering cross section is found to be
$
1.0\times10^{-52}({\rm cm^2})\lesssim\sigma_{SI}\lesssim 1.0\times10^{-50}({\rm cm^2})\ 
{\rm with}\ 1\ {\rm GeV}\lesssim {m_\chi}\lesssim 10\ {\rm GeV},
$
It implies that future experiments could be tested.  
On the other hand, in case of the bosonic DM,  the elastic scattering cross section is found to be
$
5.0\times10^{-41}({\rm cm^2})\lesssim\sigma_{SI}\lesssim 5.0\times10^{-40}({\rm cm^2})\ 
{\rm with}\ 1\ {\rm GeV}\lesssim {m_\chi}\lesssim 10\ {\rm GeV}.
$
It tells us that the range of the
cross section is consistent with  CRESSTII~\cite{cresst}, CoGeNT~\cite{cogent}, and DAMA~\cite{dama}.

\section*{Acknowledgments}
Author thanks to Dr Yuji Kajiyama and Takashi Toma for fruitful discussion.


\begin{thebibliography}{99}
\bibitem{Hehn:2012kz} 
  D.~Hehn and A.~Ibarra,
  arXiv:1208.3162 [hep-ph].

\bibitem{Ma:2006km} 
  E.~Ma,
  Phys.\ Rev.\ D {\bf 73}, 077301 (2006)
  [hep-ph/0601225].


\bibitem{Ma:1997nq} 
  E.~Ma,
  Phys.\ Lett.\ B {\bf 433}, 74 (1998)
  [hep-ph/9709474].

\bibitem{Ma:1998dp} 
  E.~Ma and D.~P.~Roy,
  Phys.\ Rev.\ D {\bf 58}, 095005 (1998)
  [hep-ph/9806210].

\bibitem{Ma:1998dr} 
  E.~Ma and U.~Sarkar,
  Phys.\ Lett.\ B {\bf 439}, 95 (1998)
  [hep-ph/9807307].

\bibitem{carena} M. Carena, A. Daleo, B. A. Dobrescu and T. M. P.
Tait, Phys. Rev. D {\bf 70}, 093009(2004)

\bibitem{t2k} T2K Collaboration: K. Abe {\it et al.}, Phys. Rev. Lett.
{\bf 107}, 041801 (2011).
\bibitem{daya} Daya Bay Collaboration: F. P. An {\it et al.},
arXiv:1203.1669 [hep-ex].
\bibitem{reno} RENO Collaboration: J. K. Ahn {\it et al.},
arXiv:1204.0626 [hep-ex].


\bibitem{Okada:2012np} 
  H.~Okada and T.~Toma,
  Phys.\ Rev.\ D {\bf 86}, 033011 (2012)
  [arXiv:1207.0864 [hep-ph]].

\bibitem{wmap}
  E.~Komatsu {\it et al.}  [WMAP Collaboration],
  Astrophys.\ J.\ Suppl.\  {\bf 192}, 18 (2011)
  [arXiv:1001.4538 [astro-ph.CO]].


\bibitem{d6-rad} 
  Y.~Kajiyama, J.~Kubo and H.~Okada,
  Phys.\ Rev.\ D {\bf 75}, 033001 (2007)
  [hep-ph/0610072].

\bibitem{pdg}
  K.~Nakamura {\it et al.}  [Particle Data Group],
  J.\ Phys.\ G {\bf 37}, 075021 (2010).

\bibitem{Adriani:2010rc} 
  O.~Adriani {\it et al.}  [PAMELA Collaboration],
  Phys.\ Rev.\ Lett.\  {\bf 105}, 121101 (2010)
  [arXiv:1007.0821 [astro-ph.HE]].


\bibitem{Okada:2012cc} 
  H.~Okada and T.~Toma,
  Phys.\ Lett.\ B {\bf 713}, 264 (2012)
  [arXiv:1203.3116 [hep-ph]].

\bibitem{Evoli:2011id} 
  C.~Evoli, I.~Cholis, D.~Grasso, L.~Maccione and P.~Ullio,
  Phys.\ Rev.\ D {\bf 85}, 123511 (2012)
  [arXiv:1108.0664 [astro-ph.HE]].

\bibitem{Dittmaier:2012vm} 
  S.~Dittmaier, S.~Dittmaier, C.~Mariotti, G.~Passarino, R.~Tanaka, S.~Alekhin, J.~Alwall and E.~A.~Bagnaschi {\it et al.},
  arXiv:1201.3084 [hep-ph]

\bibitem{Giardino:2012dp} 
  P.~P.~Giardino, K.~Kannike, M.~Raidal and A.~Strumia,
  Phys.\ Lett.\ B {\bf 718}, 469 (2012)
  [arXiv:1207.1347 [hep-ph]].


\bibitem{Higgsd1}
  S.~Chatrchyan {\it et al.}  [CMS Collaboration],
  Phys.\ Lett.\ B {\bf 716} (2012) 30
  [arXiv:1207.7235 [hep-ex]].

\bibitem{Higgsd2}
  G.~Aad {\it et al.}  [ATLAS Collaboration],
  Phys.\ Lett.\ B {\bf 716} (2012) 1
  [arXiv:1207.7214 [hep-ex]].

\bibitem{Kajiyama:2012xg} 
  Y.~Kajiyama, H.~Okada and T.~Toma,
  arXiv:1210.2305 [hep-ph].

\bibitem{higgsdecay}
https://twiki.cern.ch/twiki/bin/view/LHCPhysics/CERNYellowReportPageBR2.

\bibitem{cresst} 
  G.~Angloher, M.~Bauer, I.~Bavykina, A.~Bento, C.~Bucci, C.~Ciemniak,
	G.~Deuter and F.~von Feilitzsch {\it et al.},
  arXiv:1109.0702 [astro-ph.CO].

\bibitem{cogent}
  C.~E.~Aalseth {\it et al.}  [CoGeNT collaboration],
  Phys.\ Rev.\ Lett.\  {\bf 106}, 131301 (2011)
  [arXiv:1002.4703 [astro-ph.CO]].


\bibitem{dama}
  R.~Bernabei {\it et al.},
  Eur.\ Phys.\ J.\  C {\bf 67}, 39 (2010)
  [arXiv:1002.1028 [astro-ph.GA]].












\end{thebibliography}
\end{document}